# High piezoelectric sensitivity and hydrostatic figures of merit in unidirectional porous ferroelectric ceramics fabricated by freeze casting

Yan Zhang, James Roscow, Mengying Xie, Chris Bowen

*Department of Mechanical Engineering, University of Bath, BA2 7AY, United Kingdom*

High performance lead zirconate titanate (PZT) ceramics with aligned porosity for sensing applications were fabricated by an ice-templating method. To demonstrate the enhanced properties of these materials and their potential for sensor and hydrophone applications, the piezoelectric voltage constants ($g_{33}$ and $g_{31}$), hydrostatic parameters ($d_h$, $g_h$, $-d_{33}/d_{31}$, $d_h \cdot g_h$ and $d_h \cdot g_h/tan\delta$) and AC conductivity as a function of the porosity in directions both parallel and perpendicular to the freezing temperature gradient were studied. As the porosity level was increased, PZT poled parallel to the freezing direction exhibited the highest $d_h$, $-d_{33}/d_{31}$ and figures of merit $d_h \cdot g_h$, $d_h \cdot g_h/tan\delta$ compared to the dense and PZT poled perpendicular to the freezing direction. The $g_h$, $g_{33}$ and $g_{31}$ coefficients were highest for the PZT poled perpendicular to the freezing direction; the $g_h$ was 150% to 850% times higher than dense PZT, and was attributed to the high piezoelectric activity and reduced permittivity in this orientation. This work demonstrates that piezoelectric ceramics produced with aligned pores by freeze casting are a promising candidate for a range of sensor applications and the polarisation orientation relative to the freezing direction can be used to tailor the microstructure and optimise sensitivity for sensor and hydrostatic transducer applications.

**Introduction**

Piezoelectric materials represent a popular class of active materials used in many areas[1-3], such as SONAR applications, vibration energy harvesting, structural health monitoring and non-destructive evaluation. For uniaxial sensing applications, the piezoelectric voltage constants $g_{33}$ and $g_{31}$ are important parameters since they represent the electric field produced per unit stress, and are of interest for accelerometers, force, pressure and acoustic sensors. Hydrophones that operate under hydrostatic conditions are also an important category of piezoelectric transducers, which are employed to detect acoustic signals in water by converting the mechanical vibrations of low frequency acoustic waves into an electrical signal[4]. For such applications, the important parameters are the hydrostatic charge ($d_h$) coefficient, voltage ($g_h$) coefficient, and hydrostatic figure of merit ($FoM_1=d_h \cdot g_h$), which define the actuating (transmit) capability of the material, sensitivity of the hydrophone, and the suitability for underwater sonar applications, respectively[5, 6]. At frequencies far below the resonance frequency, energy dissipation is mainly dominated by the dielectric loss (*tan δ*), thus an alternative figure of merit of $FoM_2=d_h \cdot g_h/tan\delta$ has also been proposed[7, 8].

The hydrostatic figures of merit can be calculated from measured piezoelectric and dielectric properties as follows: $d_h=d_{33}+2d_{31}$, $g_h=d_h/\varepsilon_{33}^T\varepsilon_0$, where, $d_{33}$ and $d_{31}$ are the longitudinal and transverse piezoelectric charge coefficients, $\varepsilon_{33}^T$ is the relative permittivity at constant stress and $\varepsilon_0$ is the permittivity of the free space. These equations indicate that the important requirements for improved hydrostatic performance are a high hydrostatic charge coefficient ($d_h$) and low permittivity ($\varepsilon_{33}^T$). For many dense ferroelectric ceramic materials $d_{33} \approx -2d_{31}$ which leads to a low $d_h$, and when combined with the high permittivity of dense ferroelectrics this leads to dense materials exhibiting a low $d_h$, $g_h$, and $d_h \cdot g_h$, thereby limiting their performance as transducers under hydrostatic conditions. For uniaxial sensor applications, the combination of a high piezoelectric charge coefficient and low permittivity is also beneficial since $g_{33}=d_{33}/\varepsilon_{33}^T\varepsilon_0$ and $g_{31}=d_{31}/\varepsilon_{33}^T\varepsilon_0$.

To date, a number of researchers have made significant efforts to reduce the permittivity of the sensor material by introducing porosity into the dense material[9-12]. However, the compromise between the mechanical strength and the volume fraction and type of porosity remains a limiting issue, especially for those porous ceramics with randomly distributed porosity, which is typically achieved by a traditional processing route of adding a polymeric pore-forming agent that burns out during solid-state sintering. Furthermore, the production of aligned piezoelectric structures has recently attracted considerable interest due to their ability to reduce the resistance of piezoelectric domain motion under an applied the electric field, where alignment has been explored using nanowires[13, 14], nanofibers[15], and nanopores[16].

In this paper, we exploit the inspiration drawn from the high strength of natural nacre with a layered microstructure[17], and the ability of freeze casting, also called ice-templating[18-20] as an effective way to mimic the structure of nacre by building an oriented ceramic structure within a unidirectional temperature gradient, with the realisation of anisotropic properties in directions parallel and perpendicular to the temperature gradient/pore channel. To date, there have been a number of studies on freeze-cast piezoelectric ceramics that have utilised camphene-based [21-23], tert-butyl alcohol (TBA)-based [24-34] suspensions to achieve 3-1, 3-2 and 3-3 connectivity piezoceramic-based composites, and water-based [34-40] suspensions for 2-2 connectivity composites. The majority of the investigations above were focused on the properties of the freeze-cast parallel to the temperature gradient/pore channel, and there is little work on comparing the piezoelectric properties of freeze-cast ceramics both parallel and perpendicular to the temperature gradient/pore channel.

TBA-based $0.94Bi_{0.5}Na_{0.5}TiO_3-0.06BaTiO_3$ [30] and $K_{0.5}Na_{0.5}NbO_3$ [31] suspensions were used to achieve an aligned porous ceramic by freeze casting, whose piezoelectric constants ($d_{33}$ and $g_{33}$) and strain were examined, with differences between parallel and perpendicular orientations while the hydrostatic properties were evaluated in freeze casting camphene-based lead zirconate titanate-lead zinc niobate suspension[22]. However, camphene is potentially flammable and has been demonstrated to exhibit higher toxicity than water [41], while TBA is not only a flammable, toxic and carcinogenic substance[42, 43], but is also an emerging environmental contaminant[44]. Moreover, laminated 2-2 connectivity based piezoelectric and pyroelectric composites are of interest due to their simple architecture and superior actuation and sensing abilities[45-47]. Therefore, using water as the solvent in the freeze casting process would be a more environmentally friendly choice with lower processing cost. In addition, our previous work[48, 49] has demonstrated that porous ceramics with an aligned porous structure formed by freeze casting water-based suspension exhibited an improved mechanical strength compared to traditional porous ceramic and also lead to a significantly reduced permittivity and heat capacity compared to that of the dense material. Since the aligned structure maintained high pyroelectric properties parallel to the freezing direction a higher charging voltage and energy was achieved when charging a capacitor via the pyroelectric effect for energy harvesting applications. However, apart from our previous research on energy harvesting applications [48], there have been no reports on the piezoelectric properties by freeze casting water-based piezoelectric suspensions and their assessment in both parallel and perpendicular directions for sensor and hydrostatic applications. Therefore, this paper provides a first insight into the properties of porous piezoelectric ceramic with aligned porosity by water based freeze casting for sensor and hydrostatic applications. The piezoelectric voltage constants ($g_{33}$ and $g_{31}$), hydrostatic parameters $(d_h, g_h, -d_{33}/d_{31}, d_h \cdot g_h$ and $d_h \cdot g_h/tan\delta)$ and AC conductivity are assessed when the material was poled parallel and perpendicular directions to the freezing temperature gradient using a simple water-based suspension. Significant benefits will be demonstrated for many of the properties compared to the monolithic (dense) material, depending on the polarisation direction.

**Methods**

The raw materials used for water-based freeze casting process to fabricate the aligned porous PZT were reported in our previous work on energy harvesting[48]. Figure 1 shows images of the PZT powders before and after ball milling used in this work. It can be seen that the original as-received PZT powders exhibited a sphere-like morphology due to the spray drying processing, whose agglomerate particle diameter was approximately 40 μm, as shown in Fig. 1(A). According to our previous research[50, 51], the ceramic particle size has a strong influence on the rheological properties of the suspension, which is crucial not only for the stability of the suspension for pore preparation by freeze casting, but also the pore size of the porous channel and the final properties. Normally, the particle size of the ceramic[52] in the suspension should range from submicron to less than 3 μm in order to exhibit suitable stability and viscosity for freeze casting[53-56]. Therefore, in order to improve the rheological properties of the suspension for achieving the defect-free ceramic layer by freeze casting, a high-speed ball milling treatment in ethanol was utilised, and the particle size of the PZT powders with uniform shape were reduced to approximately 1 μm on average in Fig. 1(B).

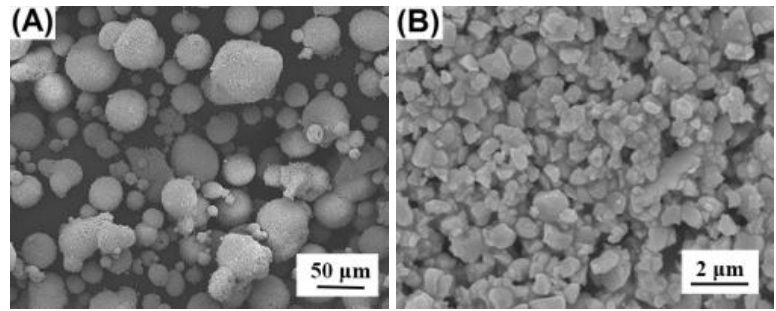

Figure 1 SEM images of PZT powders (A) without ball milling ('as-received'), (B) after ball milling for 48h.

A schematic of the freeze casting process is shown in Fig. 2. The PZT suspensions with the solid load levels of 67, 58, 48.5, 35.5, 22.5 vol.% were ball-milled for 24 h in zirconia media to generate homogeneous and fine suspensions. The prepared suspensions were de-aired (Fig. 2 (A)) before casting into a cylindrical polydimethylsiloxane (PDMS) mould with four cylinder shaped chambers (shown in Fig. S1) which was transported to a conducting cold plate in a liquid nitrogen container for the freeze casting process (Fig. 2 (B)); in this figure the freezing direction is from the base and upwards which leads to the structure shown in the image. After freezing, the frozen bodies were demoulded and freeze-dried in a vacuum at -50 °C to sublimate the ice crystal (Fig. 2 (C)) to form a 2-2-type connectivity with the PZT material aligned in the freezing direction, but more randomly orientated perpendicular to the freezing direction. To compare with the freeze cast ceramics, dense PZT pellets were formed with the initial diameter of 10 mm and thickness of 1.2 mm by uniaxial hydraulic pressing. Finally, the porous (freeze cast) and dense (pressed) green bodies were sintered at 1250 °C for 2 h under a PbO-rich after organic additives burnt out at 600°C for 3 h. Each porous PZT cylinder was poled parallel and perpendicular to the freezing direction, denoted as $PZT_\parallel$ and $PZT\perp$, respectively for the following microstructure and assessment of piezoelectric properties. A schematic showing the freezing, cutting, poling directions and SEM viewing orientation are also summarised in Fig. S2.

The microstructures of the powders and sliced PZT ceramics were examined by a scanning electron microscopy (SEM, JSM6480LV, Tokyo, Japan). The apparent porosity was derived from the density data obtained by the Archimedes method with the error of ±1.5 vol.%. Each porosity was labelled with the integer value. The remnant polarization and coercive field of the ceramics were measured using a Radiant RT66B-HVi Ferroelectric Test System on initially unpoled materials. The longitudinal piezoelectric strain coefficient ($d_{33}$) and the transverse piezoelectric strain coefficient ($d_{31}$) were measured using a Berlincourt Piezometer (PM25, Take Control, UK) after corona poling by applying a DC voltage of 14 kV for 15 min at 120 °C. The AC conductivity, $\sigma$, of the sintered ceramics were carried out from 0.1Hz to 1MHz at room temperature using an impedance analyzer (Solartron 1260, Hampshire, UK) and calculated from equation (1)[57],

$$\sigma = \frac{Z'}{Z'^2 + Z''^2} \cdot \frac{t}{A} \tag{1}$$

where $Z'$ and $Z''$ are the real and imaginary parts of the impedance, $A$ is the area of the sample and $t$ is the sample thickness.

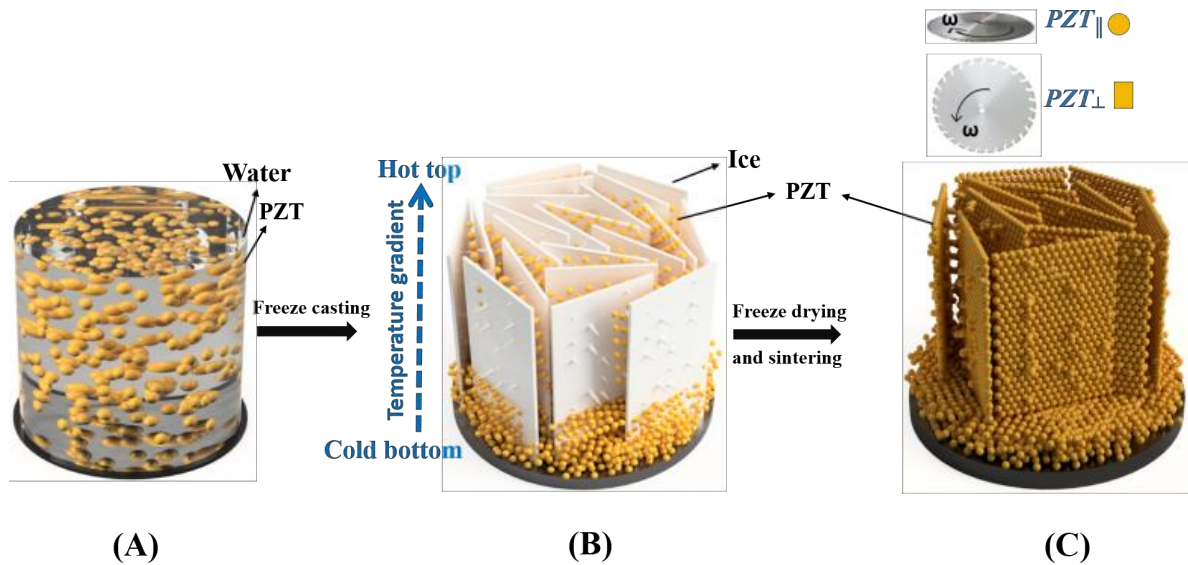

Figure 2 Schematic of porous PZT preparation by freeze casting. (A) water-based PZT suspension in each cylindrical chamber of the mould, (B) freezing of the water from the cold base, (C) freeze drying leading to a layered PZT structure. The freezing direction is indicated (form bottom to top) along with the polarisation direction for samples poled along freezing direction ($PZT_\parallel$) and perpendicular to freezing direction ($PZT\perp$).

**Results and discussion**

Figure 3(A) and (B) show SEM micrographs of the porous PZT poled parallel and perpendicular to the freezing direction, respectively. A dense lamellar ceramic wall can be seen in Fig. S3, which is desirable for high piezoelectric activity. From the macro-scale point of view, the microstructure on the top face as shown in Fig. 3(A) is equivalent to looking from above of Fig. 1(C), compared to the side face in Fig. 3(B) which is equivalent to looking from the side of Fig. 1(C). Due to the random nature of ice nucleation in the single vertical temperature gradient condition [58], there were multiple orientations on the top face of $PZT_\parallel$ as shown in the Fig. 3(A), while clear alignment of the PZT material with good parallelism of the lamellar ceramic layer can be readily observed in $PZT\perp$, although part of the lamellar layers were covered by their adjacent layers in the $PZT\perp$, as shown in the Fig. 3(B).

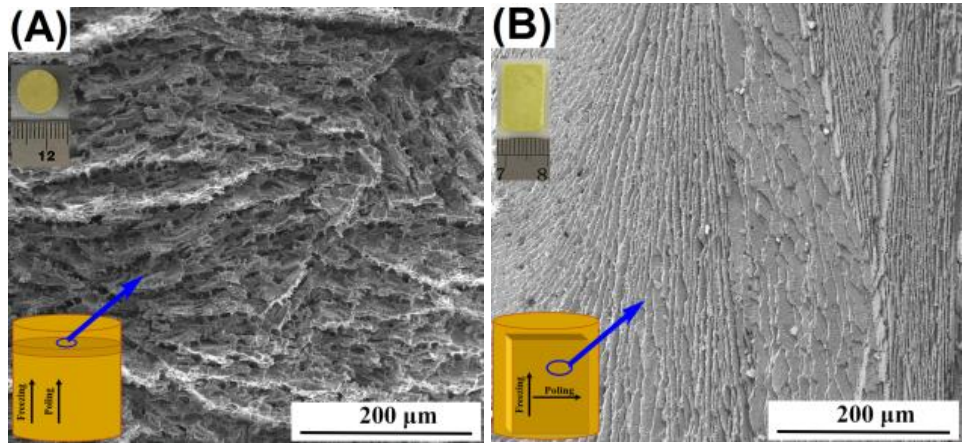

Figure 3 SEM micrographs of porous (A) $PZT_\parallel$ and (B) $PZT_\perp$.

Figure 4 (A-E) and (F-J) show SEM micrographs of the microstructure parallel ($PZT_\parallel$) and perpendicular ($PZT_\perp$) to the freezing direction as the porosity fraction is increased from 20-60 vol.%, respectively. The specific porosities of 20-60 vol.% were achieved based on the results from the initial experiments that employed a ranges of solid load levels, shown in Fig. S4. With an increase of porosity, the lamellar pore size decreased while the number of the dendrites decreased accordingly in both the $PZT_\parallel$ and $PZT_\perp$ materials. In the 20 vol.% PZT, a large quantity of the dendrites on the lamellar surface can be found, and most of the dendrites were connected with the adjacent ceramic layer, as shown in Fig. 4 (A) and (F). These additional dendrites are beneficial not only for piezoelectric phase connection, but also for the improvement of mechanical strength. It can be seen that when the porosity reached 60 vol.%, the surface of the lamellar layers became relatively smooth with ceramic links on the edge of the layers, as shown in Fig. 4(E) and (J). In the freeze casting process, a low freezing rate can provide a longer period for ice growth, which is the replica of the lamellar pore in the SEM images, while a low solid loading can provide a low viscosity which facilitates ceramic particle rearrangement. Both of the above conditions can lead to a large lamellar pore size and the high porosity materials exhibiting a smaller number of dendrites.

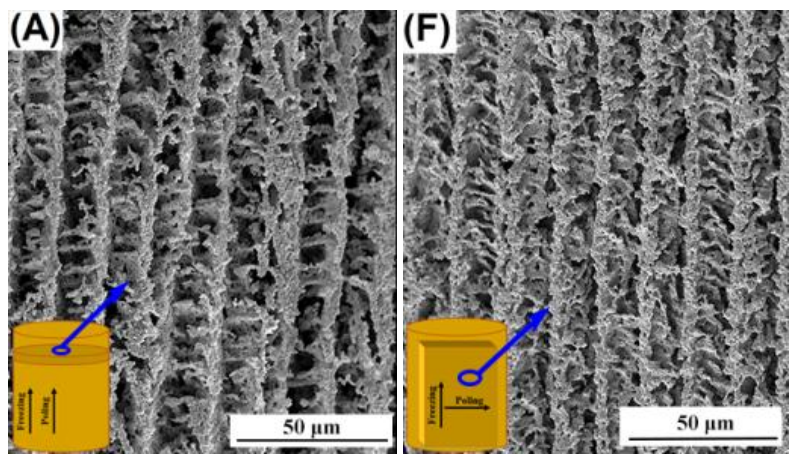

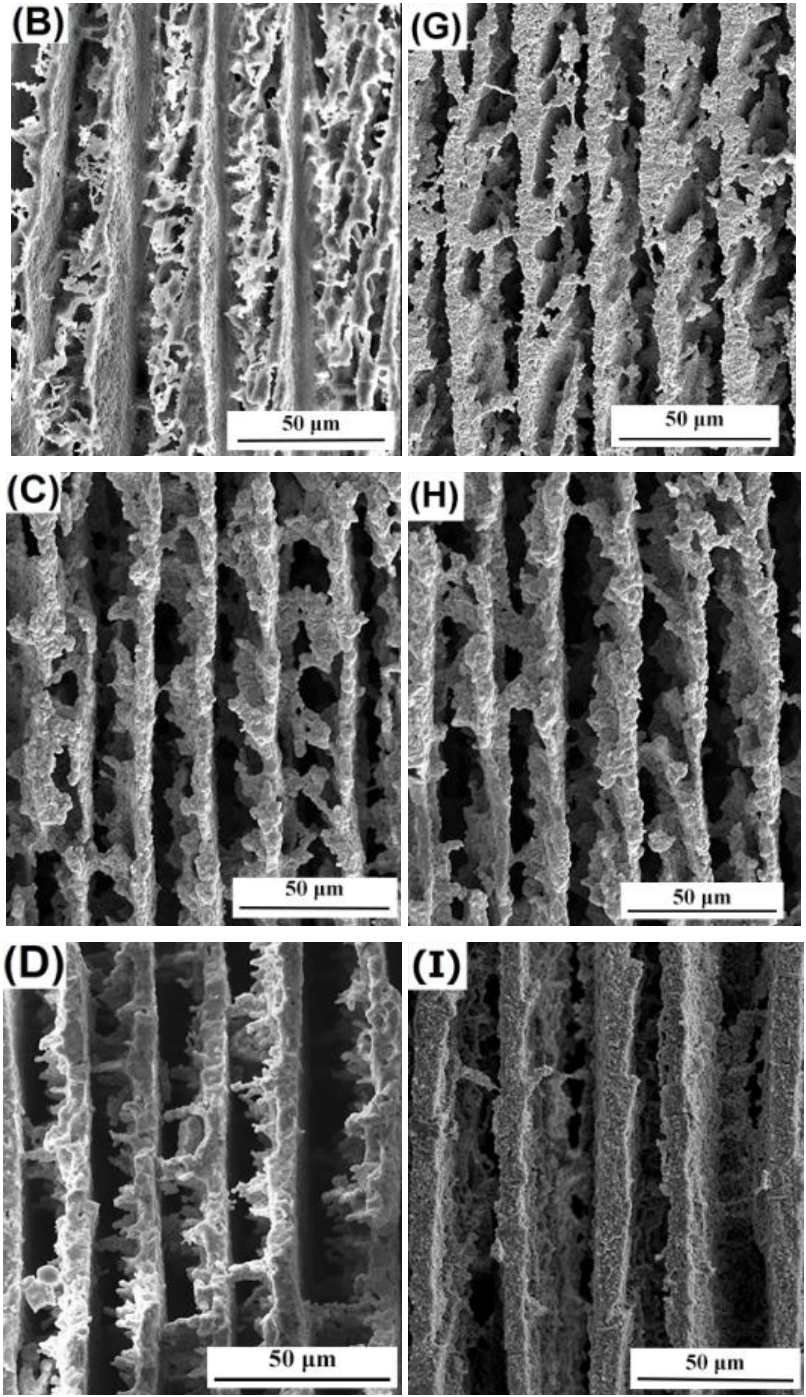

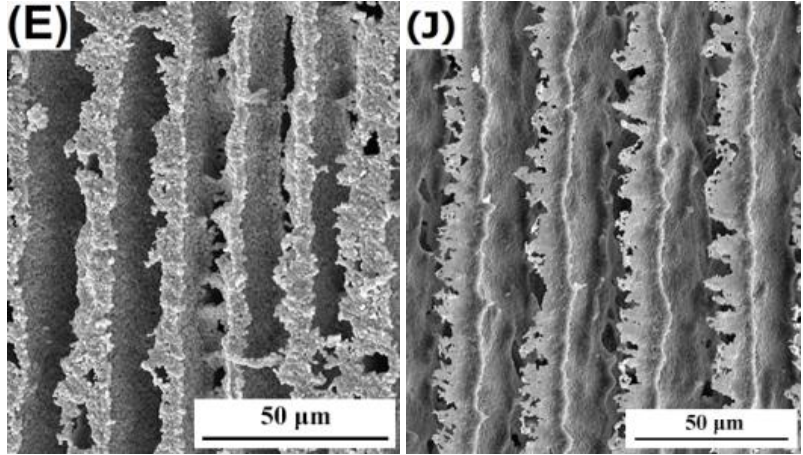

Figure 4 SEM images of porous $PZT_\parallel$ with the porosity of (A) 20, (B) 30, (C) 40, (D) 50, and (E) 60 vol.% and $PZT\perp$ with the porosity of (F) 20, (G) 30, (H) 40, (I) 50, and (J) 60 vol.%.

Figure 5 shows the remnant polarisation ($P_r$) and coercive field ($E_c$) of the initially unpoled porous PZT as a function of the porosity ranging from 20 to 60 vol.%, respectively. It can be observed that the $P_r$ of both $PZT_\parallel$ and $PZT\perp$ decreased with an increase of porosity, which were 2.1-5.5 and 3.1-10.0 times lower than the dense PZT, respectively. The dense material is characterised as having 4 vol.% porosity, with the density of 7.2 g/cm$^3$ based on the theoretical density of 7.5 g/cm$^3$ from the datasheet provided by the supplier. The reduction in $P_r$ is likely to be due to the reduced amount of polarised material and the inhomogeneous electric field distribution in the porous materials as a result of the contrast in permittivity between the high-permittivity PZT and low-permittivity air. The decrease of $P_r$ is also associated with its connectivity (Fig. 3A), and leads to a decrease of piezoelectric properties[48], such as $d_{33}$ and $d_{31}$. The decrease in $P_r$ with porosity is more rapid than predicted by $P_r^{porous}=P_r^{dense}\times(1-p)$ where $p$ is porosity [59]; this is due to the porosity also restricting polarisation of the ceramic since the electric field concentrates in the lower permittivity pore space. The $PZT_\parallel$ material exhibited a 1.5-1.8 times higher $P_r$ compared to the $PZT\perp$ due to the improved connectivity of piezo-active material along the freezing direction and therefore the lower fraction of unpoled areas in $PZT_\parallel$[48, 60]. The $E_c$ values of both $PZT_\parallel$ and $PZT\perp$ increased as the porosity increased, see Fig. 5B, and the $PZT\perp$ exhibited the highest $E_c$ values in all ranges of porosity. The increase in $E_c$ with porosity is due to the concentration of the applied electric field in the low permittivity pore space, leading to higher applied electric field being required to achieve domain switching in the higher permittivity ferroelectric phase. For the same reason, a higher $E_c$ is observed for the $PZT\perp$ material as there is a reduced connection of ferroelectric along this poling direction as a result of the overlapped lamellar layers, see Fig. 3(D), resulting in a reduced piezoelectric response[61] and a higher $E_c$ in $PZT\perp$. The inhomogeneous electric field due to the presence of pores is also reflected in the reduced rectangularity ($P_{remnant}/P_{saturation}$) of the materials as the porosity level increases, as seen in Fig. 5(C).

Interestingly, the $PZT_{\parallel}$ with the porosity of 20 vol.% exhibited the lowest $E_c$ of 7.7 kV/cm compared to both the dense (8.7 kV/cm), and all the $PZT\perp$ materials, demonstrating easier switching of ferroelectric domains with applied electric field. The presence of a small amount of porosity (~20 vol.%) may initially provide a state of reduced internal stress or restriction of domain motion, while at higher porosity levels the applied electric field concentrates in the pore volume and leads to higher applied electric fields being required to provide domain switching. Therefore, although the existence of porosity can facilitate the switching of the domain walls to some extent[62, 63], the increased electric field concentration[60] in the low permittivity pore space leads to a higher $E_c$ which reached to 8.9 μC/cm$^2$ compared to the dense with the $E_c$ of 8.7 μC/cm$^2$ when the porosity was higher than 50 vol.%.

In addition, due to randomness of the lamellar pores orientations perpendicular to the freezing direction in Fig. 3 (A) and Fig. 2 (C), the angle of the pore in $PZT\perp$ ranged from >0˚to 90˚, while $PZT_{\parallel}$ had an angle of ~0 ˚along the poling direction; see Fig. S5. This leads to a better level of poling of $PZT_{\parallel}$ [64] compared to $PZT\perp$, leading to a higher polarisation for $PZT_{\parallel}$ [48, 64], as seen in Fig. 5(A).

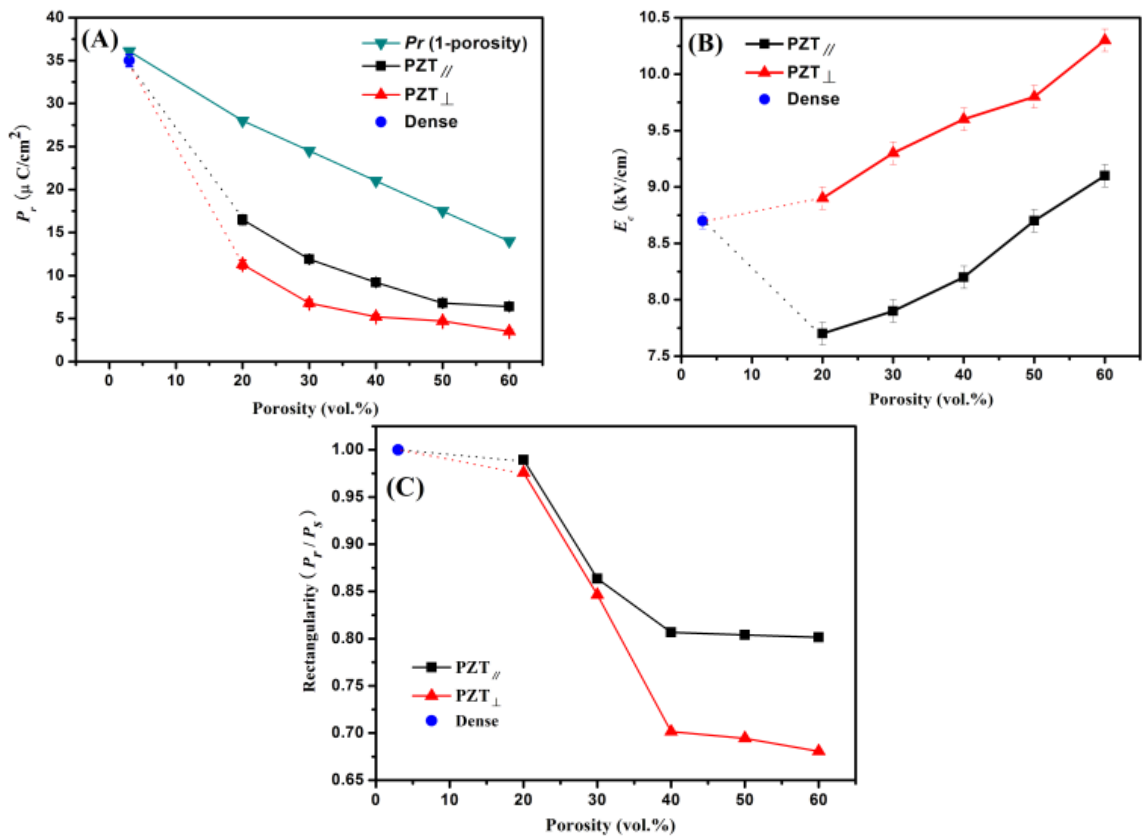

Figure 5 (A) Remnant polarisation ($P_r$), along with estimation based on $P_r^{porous}=P_r^{dense}\times(1\text{-}p)$, (B) coercive field ($E_c$) of the porous freeze-cast PZT, and (C) rectangularity ratio of with $P_r / P_s$ various porosities. The dense material is also shown.

Figure 6 (A-F) show the anisotropy factor[65] $-d_{33}/d_{31}$, hydrostatic charge ($d_h$), voltage coefficient ($g_h$), piezoelectric voltage coefficients ($g_{33}$ and $g_{31}$), relative permittivity ($\varepsilon_{33}^T$) and figures of merit ($d_h g_h$ and $d_h g_h/tan\delta$) of the porous PZT as a function of the porosity ranging from 20 to 60 vol.% and dense PZTs. It can be seen from Fig. 6(A) that $PZT\bot$ exhibited a lower $-d_{33}/d_{31}$ than the dense PZT at all porosities, while porous $PZT\parallel$ exhibited a higher $-d_{33}/d_{31}$ and therefore higher anisotropy than the dense PZT. The $-d_{33}/d_{31}$ $PZT\parallel$ increased with increasing porosity and was 1.2-2.0 times higher than that of $PZT\bot$. The increase in piezoelectric anisotropy for $PZT\parallel$ is advantageous since it leads to higher $d_h$ values that were determined from the $d_{33}$ and $d_{31,}$ as shown in Fig. 6(B). It can be seen the $PZT\bot$ exhibits a gradual reduction in $d_h$ with increasing porosity, due to a reduced $-d_{33}/d_{31}$ while the $PZT\parallel$ exhibits an increase in $d_h$ with increasing porosity. It should be also noted in Fig. 6(B) that the $d_h$ of the $PZT\parallel$ was higher than the dense PZT when the porosity exceeded 40 vol.%. The increase in $d_h$ for the porous material compared to the dense material is relatively modest, this is due to the fact that the dense material already has a relatively high degree of anisotropy with a $-d_{33}/d_{31}$ of ~3 (see Figure 6A); typically $-d_{33}/d_{31}$ is close to 2 for dense PZT based materials.

While the $PZT\bot$ materials exhibited relatively poor $d_h$ values it exhibits advantageous $g_h$ values that are 1.2 to 2.1 times and 1.5-8.5 higher than $PZT\parallel$ and monolithic dense PZT, respectively; see Fig. 6(B), i.e. $40.1\times10^{-3}$ $Vm/N$ for $PZT\parallel$, and $83.5\times10^{-3}$ $Vm/N$ for $PZT\bot$ both at 60 vol.% porosity compared to that of dense PZT ($9.0\times10^{-3}$ $Vm/N$). This was due to the reduced relative permittivity, shown in Fig. 6(C) and (D), of the $PZT\bot$ ($\varepsilon_{33}^T$ ~ 380 to 16) compared to $PZT\parallel$ ($\varepsilon_{33}^T$ ~ 1407 to 581) and the dense material ($\varepsilon_{33}^T$ = 2158); at 1 kHz from the inset of Fig. 6(C) and (D). The $g_h$ value of the $PZT\parallel$ was also 1.3-4.1 times higher than that of the dense PZT (see Fig. 6(B)) since the dense material exhibited a much higher permittivity, Fig. 6(C).

A similar trend is observed on examination of the piezoelectric voltage coefficients ($g_{33}$ and $g_{31}$) in Fig. 6 (E) where both $g_{33}$ and $g_{31}$ of $PZT\bot$ were 1.8-5.2 and 2.0-10.0 times higher than $PZT\parallel$, and also 2.3-12.5 and 2.5-14.7 times higher than dense PZT. This indicates that $PZT\bot$ is attractive as a piezoelectric force/pressure sensor. Fig. 6(F) shows that both the $d_h \cdot g_h$ and $d_h \cdot g_h/tan\delta$ figures of merit for the $PZT\parallel$ increased with increasing porosity, and were much larger than both the dense material and $PZT\bot$. Although a reduced inter-connection of the piezoelectric phase was observed with an increase of porosity (see Fig. 4), the high degree of alignment of the pore channel along the poling direction can compensate for the reduction in interconnectivity, especially in $PZT\parallel$. This is due to a combination of, relatively high piezoelectric activity (Fig. 5), high piezoelectric coefficients, high anisotropy and low permittivity achieved through the introduction of the porosity. These results compare favourably with previous analysis[48] that demonstrated the higher $d_{33}\cdot g_{33}$ piezoelectric and pyroelectric harvesting figures of merit (*(pyroelectric coefficient)$^2$/ permittivity·heat capacity*) in $PZT\parallel$.

The highest values of hydrostatic figures of merit were achieved for $PZT\parallel$ when the porosity increased to the maximum value of 60 vol.% in this work; this corresponded to hydrostatic figures of merit that were 2.7-10.2 and 2.0-10.9 times higher than the dense materials and $PZT\perp$, respectively, as shown in Fig. 6(D). The 60 vol.% was chosen as a maximum since for higher porosity levels the material will exhibit reduced mechanical properties, and an even higher coercive field; leading to a low $d_{33}$ and therefore a low $d_h$. In addition to the advantages of high mechanical strength reported previously[48, 49], the freeze-cast samples exhibited comparable or even higher hydrostatic figure-of-merit than most of other processing methods, especially $PZT_\parallel$ which exhibit a higher piezoelectric coefficient and lower permittivity, as shown in Table 1. It can be seen that freeze casting generally leads to high hydrostatic figures of merit compared to other methods. Very high figures of merit are reported in Table 1 for PZT-lead zirconate niobate (PZN) [22], which is due to the high porosity levels (90vol.%); although mechanical properties can be a concern at such low ceramic volume fractions.

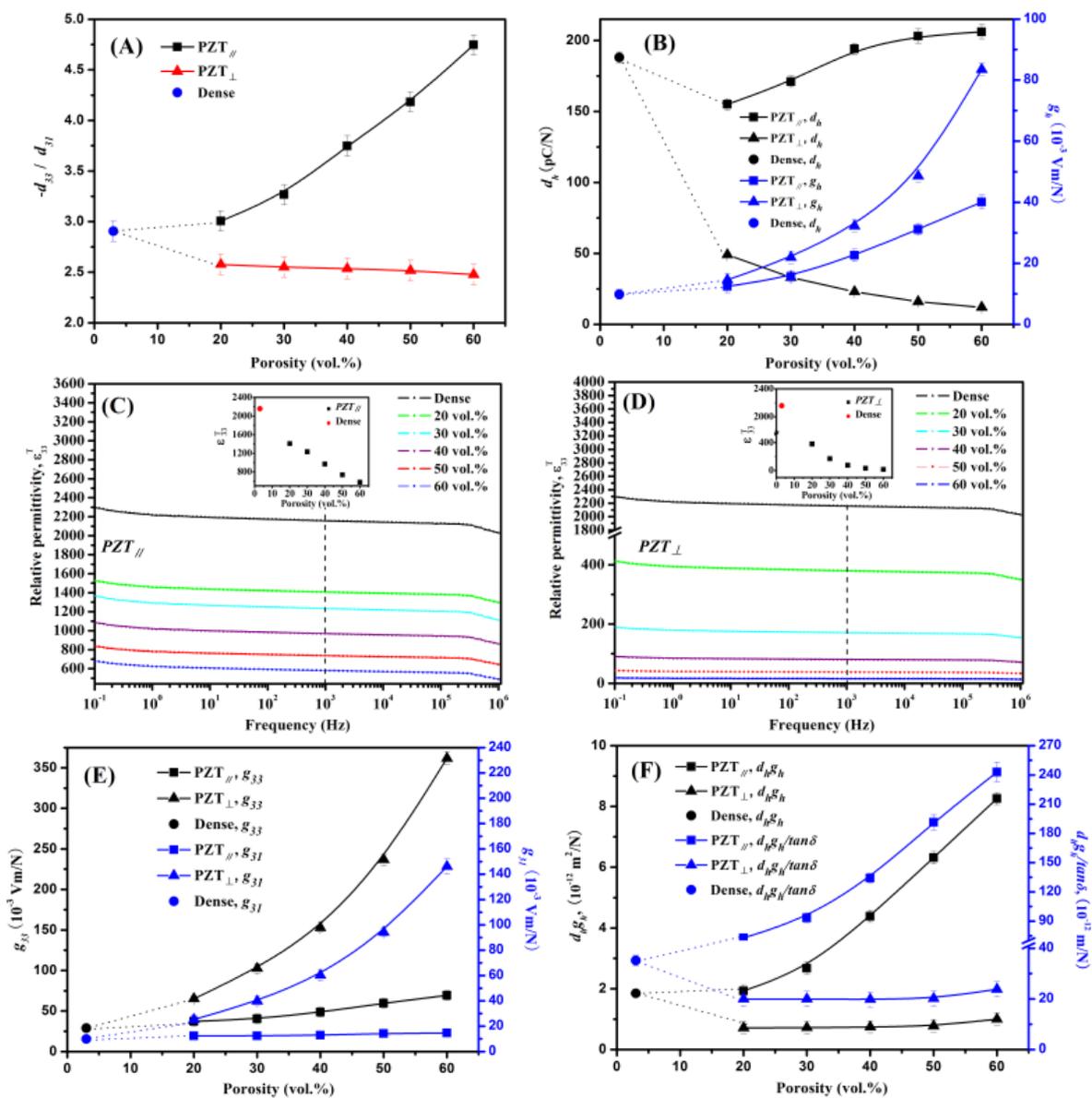

Figure 6 (A) anisotropy factor of $-d_{33}/d_{31}$, (B) hydrostatic charge ($d_h$) and voltage coefficient ($g_h$), (C) (D) relative permittivity ($\varepsilon_{33}^T$), (E) piezoelectric voltage coefficients ($g_{33}$ and $g_{31}$), and (F) hydrostatic figure of merits ($d_hg_h$ and $d_hg_h/tan\delta$) of the porous freeze-cast PZT with various porosities. The dense material is also shown.

Table 1 Comparison of hydrostatic parameters with different processing methods for lead zirconate titanate (PZT) or PZT-based composite. PZN = lead zircontate niobate.

| Production method | Composite | | Connectivity | PZT Volume % | $d_h$ (pC/N) | $g_h$ ($10^{-3}$ Vm/N) | $d_h g_h$ ($10^{-12}$ m$^2$/N) | Ref. |
|---|---|---|---|---|---|---|---|---|
| Freeze casting | PZT-air (water-based) | Parallel | 2-2 | 40 | 206 | 40.1 | 8.26 | This work |
| | | Perpendicular | | | 12 | 83.5 | 1 | |
| | PZT/PZN-air (camphene-based) | Parallel | 3-3 | ~10 | 406 | 396 | 161.01 | [22] |
| | | Pore orientation of 45° to poling direction | | | ~350 | ~330 | ~115.5 | |
| | | Perpendicular | | | 216 | 241 | 53.13 | |
| | PZT/PZN-air (camphene-based) | Parallel | 3-3 | ~48 | ~250 | ~32 | ~8 | [23] |
| | PZT-air (TBA-based) | Parallel | 1-3 | 31.3 | - | - | 9.32 | [25] |
| | PZT-air (TBA-based) | Parallel | 3-3 | ~60 | - | - | 7.6 | [33] |
| | PZT-air (TBA-based) | Parallel | 3-3 | 41.4 | 200 | - | 10.11 | [32] |
| | PZT-air (alginate/water-based) | Parallel | 3-1 | 43.22 | ~223 | - | ~5.7 | [37] |
| BURPS (Burnout of Polymer Spheres) | PZT-air | | 3-0 | 54.5 | ~35 | ~16.4 | ~0.57 | [66] |
| | | | 3-1 | 59 | - | ~42 | ~5 | [67] |
| | | | 3-3 | ~60 | 102 | 72 | 7.3 | [10] |
| | | | 3-0 | 68 | - | ~48 | 5 | [68] |
| Ionotropic gelation process | PZT-air | | 3-1 | ~44 | ~222 | - | ~5.6 | [37] |
| Pore-forming agent | PZT-epoxy | | 3-1 | 45-60 | 60 | 69 | ~4.01 | [69] |
| Polymer injection (PZT rods embedded in polymer) | PZT-polymer | | 1-3 | 40 | ~95 | ~18 | ~1.71 | [70, 71] |
| Direct-write | PZT-polymer | | 2-2 | ~30 | <190 | <0.38 | <0.72 | [72] |
| Dice-and-fill | PZT-polymer | | 1-3 | 40 | ~135 | ~37 | ~5 | [73] |
| | PZT-cement | | 2-2 | 40 | <100 | <0.2 | <0.02 | [74] |

Figure 7 (A) and (B) show the AC conductivity ($\sigma$) of the $PZT_\parallel$ and $PZT_\perp$ at frequencies ranging from 0.1 to $10^6$ Hz at room temperature respectively. It is clear from these two figures that the AC conductivity decreased with an increase of the porosity in both PZTs and were lower than that of the dense PZT at all the porosities and the frequencies; this includes the lowest frequencies where the conductivity is becoming less frequency dependent and is approaching the DC conductivity. The $PZT_\parallel$ possessed a higher AC conductivity than the $PZT_\perp$ at the same porosity and frequency, e.g. 1.3-

1.5 times at the frequency of 1 kHz, as shown in the insets in Fig. 7 (A) and (B). This is likely to be due to the high permittivity of the $PZT_\parallel$ resulting from the contribution of the dielectric phase to the overall conductivity ($\omega\varepsilon_r\varepsilon_0$)[75]. Generally, there are two well-known models for interpreting the effects of porosity on electrical conductivity, where the solid phase has a small, but finite conductivity, and the pores have a very small (almost negligible) conductivity. For a composite consisting of both PZT and air, the low frequency conductivity parallel to the poling direction $\sigma_\parallel$ (parallel connected) and PZT perpendicular to the freezing direction (series-connected) $\sigma_\perp$, which can be calculated as $\sigma_\parallel = v_{PZT}\cdot\sigma_{pzt} + v_{air}\cdot\sigma_{air}$ and $\sigma_\perp = \frac{\sigma_{PZT}\cdot\sigma_{air}}{V_{air}\cdot\sigma_{PZT}+V_{PZT}\cdot\sigma_{air}}$, where $v_{PZT}$ and $v_{air}$ are the volume fractions of PZT and air, $\sigma_{pzt}$ and $\sigma_{air}$ are the electric conductivity of the dense PZT and air[76]. Although $\sigma_{pzt} \gg \sigma_{air}$, no linear relation was found between conductivity $\sigma$ and the porosity, as shown in the insets in Fig. 7 (A) and (B), owing to the existence of the dendritic ceramic link between the adjacent ceramic layers shown in Fig. 4, which means the pore regions were a mixture of the PZT ceramic links and the air. Similar situations can be also found in the piezoelectric and pyroelectric properties[48]. Furthermore, along the electric field direction during the impedance measurement, much more effective interface areas between the active phase PZT and the passive phase air were formed in the porous $PZT_\parallel$, therefore, the better ability of the domain movement and carrier mobility were the main reasons for the higher conductivity in the porous $PZT_\parallel$, making it be a more suitable candidate for hydrostatic sensor applications.

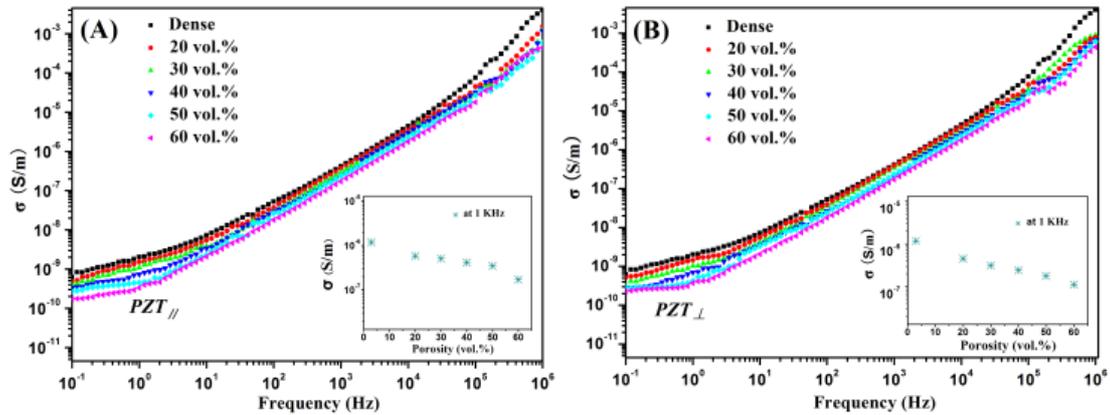

Figure 7 AC conductivity ($\sigma$) of the (A) $PZT_\parallel$ and (B) $PZT_\perp$ with various porosities. Insets in (A) and (B) are $\sigma$ values of the $PZT_\parallel$ and $PZT_\perp$ at 1 kHz as a function of porosity, respectively. The dense material is also shown.

**Conclusions**

Freeze casting using a water based suspension has been utilised and demonstrated to be an effective method to prepare high performance porous PZT for sensor applications with unidirectional pore channels over a range of 20-60 vol. % porosity. The hydrostatic and sensor properties of PZTs poled parallel ($PZT_\parallel$) and perpendicular ($PZT_\perp$) to the freezing direction were examined in detail.

Significant improvements in the hydrostatic figures of merit were observed compared to the dense monolithic material. In terms of $d_h \cdot g_h$ and $d_h \cdot g_h / tan\delta$, the $PZT_\parallel$ and $PZT_\perp$ were 2.7-10.2 and 2.0-10.9 times higher than that of the dense material. These highly attractive properties are due to their relatively high $d_{33}$, reduced $d_{31}$ and significantly reduced permittivity. While the $PZT_\parallel$ exhibited the best $d_h$, $d_h \cdot g_h$ and $d_h \cdot g_h / tan\delta$, the $PZT_\perp$ exhibited the highest $g_h$, $g_{33}$ and $g_{31}$ coefficients which was attributed to the lower permittivity of the material in this orientation. In addition, the $PZT_\parallel$ showed 1.3-1.5 times higher AC conductivity than the $PZT_\perp$ at the frequency of 1 kHz. In addition, it is shown that the properties are improved compared to piezoelectric composite materials manufactured by more complex methods. The $PZT_\perp$ exhibited the highest $g_h$, $g_{33}$ and $g_{31}$ coefficients; up to 1.5-8.5, 2.3-12.5 and 2.5-14.7 times higher compared to the dense materials; this was attributed to the low permittivity of the material in this orientation. The coercive field increased with an increase in porosity for the materials, and this was attributed to the concentration of electric field concentration in the lower permittivity pore space. This work demonstrates water-based freeze casting provides an environmentally friendly and promising route to fabrication porous piezoelectric for both uniaxial and hydrostatic sensing applications with potential for control of the coercive field of ferroelectric materials. It is also demonstrated that changing of the poling condition relative to the freezing direction can enable control and optimisation of the relevant performance figures of merit for specific applications. In addition to varying the solid load levels, the ability of the freeze casting method to adjust the pore size, the thickness of the ceramic wall and the density of the ceramic wall in the aligned pore structures by tailoring the freezing rate, freezing temperature and temperature gradient provides significant versatility for this promising processing route to create new sensor materials.

## Acknowledgment

Dr. Y. Zhang would like to acknowledge the European Union's Horizon 2020 research and innovation programme under the Marie Skłodowska-Curie Grant, Agreement No. 703950 (H2020-MSCA-IF-2015-EF-703950-HEAPPs). Prof. C. R. Bowen, Dr. M. Xie and Mr J. Roscow would like to acknowledge the funding from the European Research Council under the European Union's Seventh Framework Programme (FP/2007–2013)/ERC Grant Agreement No. 320963 on Novel Energy Materials, Engineering Science and Integrated Systems (NEMESIS).